\documentclass[sigconf]{acmart}

\def\BibTeX{{\rm B\kern-.05em{\sc i\kern-.025em b}\kern-.08emT\kern-.1667em\lower.7ex\hbox{E}\kern-.125emX}}
\usepackage{url} 
\copyrightyear{2019}
\acmYear{2019}

\begin{document}

%
\title{Automatic Realistic Music Video Generation from Segments of Youtube Videos}

%
\author{Sarah Gross, Xingxing Wei, Jun Zhu}
\affiliation{Department of Computer Science and Technology, Tsinghua University}


  

%

%
\begin{abstract}
A Music Video (MV) is a video aiming at visually illustrating or extending the meaning of its background music. This paper proposes a novel method to automatically generate, from an input music track, a music video made of segments of Youtube music videos which would fit this music. The system analyzes the input music to find its genre (pop, rock, ...) and finds segmented MVs with the same genre in the database. Then, a K-Means clustering is done to group video segments by color histogram, meaning segments of MVs having the same global distribution of colors. A few clusters are randomly selected, then are assembled around music boundaries, which are moments where a significant change in the music occurs (for instance, transitioning from verse to chorus). This way, when the music changes, the video color mood changes as well.
This work aims at generating high-quality realistic MVs, which could be mistaken for man-made MVs. By asking users to identify, in a batch of music videos containing professional MVs, amateur-made MVs and generated MVs by our algorithm, we show that our algorithm gives satisfying results, as 45\% of generated videos are mistaken for professional MVs and 21.6\% are mistaken for amateur-made MVs. More information can be found in the project website: http://ml.cs.tsinghua.edu.cn/\url{~}sarah/

\end{abstract}

%
%


%

%

%
\maketitle

\section{Introduction}
With the progress of Artificial Intelligence, researchers tried to question the understanding of machine produced art. Many works have been done on text generation such as creating poems \cite{Yi17}, on image generation by generating images in the style of Van Gogh\cite{Gatys15}, or even on music generation with the creation of polyphonic chorales in the style of Bach \cite{HadjeresP16}. 

However, videos are the forgotten ugly duckling of media generation research. Best results in video generation \cite{Vondrick16} \cite{Pan17} \cite{Li17} so far can only achieve very limited samples,  such as GIFs of maximum $64 \times 64$ pixels, still lacking coherence in the structure hence realism, and only in the very specific categories they were trained on (\textit{e.g.} bouncing digit, kite surfing, baby ...). Therefore generating video from scratch is currently a very difficult task unable to produce results comparable to real videos.

Yet, videos are nowadays the most popular media on the net. According to Cisco Visual Networking Index Forecast \cite{onlineCisco}, IP video traffic represented 75\% of overall IP traffic in 2017 and will grow up to 82\% in 2022. Around 100 million hours of video are watched everyday on Facebook \cite{onlineFacebookVideos}, and Youtube is the world's largest search engine after Google \cite{onlineYoutubeSearch}. Influencing power of videos is tremendous, as a video Tweet is 10x more likely to be retweeted than a photo Tweet \cite{onlineRetweet}. For that reason, companies invested dramatically in this media, increasing branded video content by 99\% on YouTube and 258\% on Facebook between 2016 and 2017 \cite{onlineBranded}.

Within this golden media, music videos strive. "Music" is the first-searched term on Youtube in 2017 \cite{onlineYoutubeSearch}. Within the top 30 most-viewed videos on Youtube, each accumulating more than 2 billion views, only two are not MVs \cite{onlineWiki}. Music videos represent great artistic value, as it combines two medias on which artistic work is already performed, music and video, to create an enhanced media. Being able to generate a music video is therefore a challenge with  the potential to raise significant interest from the public towards AI research.

A few authors previously worked on MV generation, but none of them produced MVs from a music comparable to real music videos produced for an artist. As these kinds of videos receive high interest from the Internet community, with fans creating their own alternatives of music videos when the artist do not provide one, our work would provide not only an innovative method for MV generation, but also an alternative source of music videos for musics deprived of official video clips.

From a database of music videos downloaded from Youtube-8M dataset, we extract video shots and store the color histogram for each shot. Given this configuration, our generation algorithm can create in only a hundred seconds a suitable music video for an audio file given in input. First, we identify the genre of the input music in order to use only video segments in our database originating from MVs of the same genre. Second, these video shots are grouped together based on their color histograms using K-Means algorithm to create K clusters of similar color distributions. Key changes in the song (boundaries between sections such as verse, chorus, bridge, etc.) are detected. Finally a few clusters are selected and consecutively appended between each boundary, triggering a cluster change, and thus, a major color change when a music boundary is encountered.

Our contributions are summarized as follows: 
($i$) To the best of our knowledge, we are the first to generate a realistic music video according to an input music, and propose a systematic method using a database of Youtube music videos. ($ii$) Contrary to previous works in music video generations, our selection of features (color histogram, key changes in music, genre) are all justified using sociological studies of music videos. ($iii$) Our method is tested through a user study where in more than half of cases users mistake our generated videos for human-made videos, and confirm with feedback our choice of features. ($iv$) Our algorithm provides helper functions such as music genre recognition and video inner resolution harmonization which can be used in future works.

The remainder of this paper is organized as follows. Section 2 briefly reviews the related work. Section 3 details the database configuration. Section 4 introduces the music video generation method. Experimental results are presented in Section 5, followed by conclusion in Section 6.

\section{Related work}

Earlier works in MV generation relied on assembling images fetched from the internet based on lyrics content \cite{Wu12} \cite{Cai07}, creating a video output similar to a slideshow. 
Another category of researchers decided not to base their MVs on still images, improving the output quality, however the base video was a unique video shot in amateur conditions. In order to generate artistic-looking summary videos of private events, Yool et. al. \cite{Yoon08}, Hua et. al. \cite{Hua04} and Foote et. al. \cite{Foote02} take in input a user home video in addition to the music, analyze the video, segment it and assemble it around the music accordingly, with features specific to each work. Their approach differs significantly from our work in two respects. On the one hand, they have additional constraints : since they deal with amateur, homemade video they need extra video pre-processing and analysis to discard low-quality shots, and must put in extra work to match the video with the music since the home video likely contains actions that are not especially aesthetic or contain any rhythm. On the other hand, our database is made of tens of thousands segments issued from about a thousand MVs, which therefore have different styles and looks, and feature different people. This gives us an additional constraint to assemble these segments while keeping some consistency throughout the video.

Finally, most recent works do the opposite process by taking one input video and finding the best matching music to generate a music video. Lin. et. al. \cite{Lin16} \cite{Lin17} trained a Deep Neural Network on the 65 MVs of DEAP Database to find emotional correspondence between video content and music content. By applying this trained algorithm on a video, the algorithm picks a music in a database of audio files which would "emotionally" fit it best. Though this approach is highly innovative and technically interesting, it produces unconvincing results for two reasons: first, the video transitions have no relationship at all with the music; second, emotional correspondence between the video atmosphere and music is actually not so important in professional MVs. For artistic reasons, music video creators do not focus much on the emotional feeling of the music and include many elements which have absolutely no correlation, for instance "performance" shots where we see close-ups of the band playing musical instrument. In Electronic genre musics, since the melody relies on strong beats and dancing atmosphere, one could think of always pairing them up with lively, joyful videos; however very few electronic MVs follow this rule, like Marshmello's single "Alone" which, in spite of having a lively melody, depicts the sad everyday life of a bullied high-schooler.

The current state of video generation  \cite{Vondrick16} \cite{Pan17} \cite{Li17} shows that generating an MV from scratch would be impossible. Instead, unlike previous works, we will use extracts of real MVs to generate a music video for an input music.

\section{Database constitution}
\begin{figure}[h]
  \centering
  \includegraphics[width=0.9\linewidth]{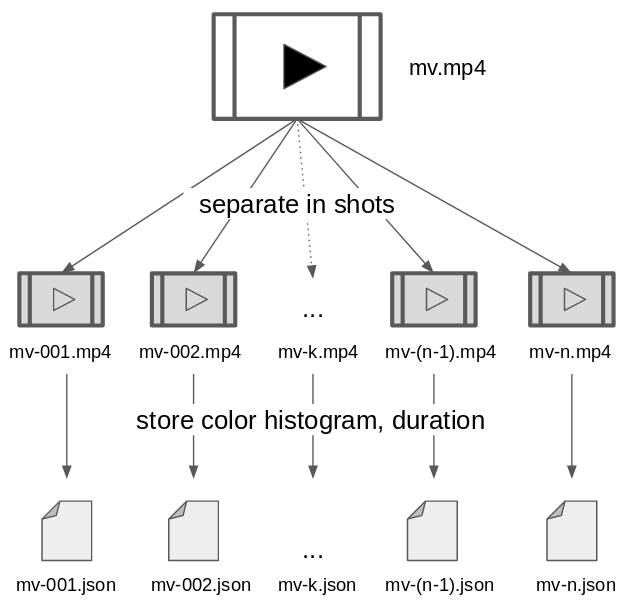}
  \caption{Database constitution process.}
  \Description{Database constitution process : 1. segmentation in shots, 2. for each shot, extract color histogram and shot duration and store it in a json file.}
  \label{fig:database}
\end{figure}

Contrary to previous works, our study relies on a very large dataset: \textbf{Youtube-8M}, released in 2017 to help the improvement of video-based research. Among these 8 million videos, "Music Video" represents the fourth most frequent entity \cite{Youtube8M}. By querying Youtube search API with the keywords "official music video", K. Choi provides the subset YouTube-music-video-5M \cite{Youtube5M}, a list of 5,119,955 video ids.
Approximately 1000 videos from YouTube-music-video-5M dataset are downloaded to test our algorithm. We extract several information from these music videos in order to prepare in advance the database, hence minimizing the number of operations required in the future by the MV generation algorithm.
We perform the process illustrated by figure~\ref{fig:database} on each video of the database :
\begin{enumerate}
    \item Separate the video into shots, also called \textit{scenes}. A scene is made of one single motion of the camera, without any cut.
    \item Calculate the color histogram for each scene.
    \item Store the color histogram array and the duration for each scene into a \verb+json+ file.
\end{enumerate}

The database is split into folders of sub-videos so that the generation algorithm can select a subset of scenes originating from different MVs, to finally concatenate them into a music video. This project \verb+github+ provides all the guidance and functions for automatically converting a database of downloaded music videos to the format described here.

\subsection{Video segmentation}

Existing shot boundary detection methods can generally be categorized \cite{Yuan07} into rule-based ones and machine learning based ones. Although Li et. al. \cite{Li09} judge machine learning methods more precise, namely because of the difficulty to find a proper threshold in rule-based methods, we decide to use video segmentation algorithm provided by Python module \verb+PySceneDetect+, relying on very simple threshold-based algorithms, due to its goods performances.

\verb+PySceneDetect+ provides two different methods in order to detect scene transition:
\begin{itemize}
    \item \textbf{Content-Aware Detection Mode} this mode looks at the difference between the pixels in each pair of adjacent frames, triggering a scene break when this difference exceeds a given threshold $t$. Zhang et. al. \cite{Zhang93} call this method \textit{Pair-wise comparison}. For a pixel of two-dimensional coordinates $(a,b)$ and a frame $i$ currently being compared with its successor, we define the binary function $DP_i(a,b)$ depending on $P_i(a,b)$ the intensity value of the pixel: 
    \begin{equation*} 
    DP_i(a,b) =
      \begin{cases}
        1       & \quad \text{if } |P_i(a,b)-P_{i+1}(a,b)|>t\\
        0  & \quad \text{otherwise}
      \end{cases}
    \end{equation*}
    
    The content-aware mode counts the number of pixels differing from one frame to its successor. If we consider a frame a dimensions $M\times N$, a segment boundary is detected if the percentage of different pixels according to the metric $DP$ is greater than a given threshold $T$:
    \begin{equation*} 
        \frac{\sum_{a,b}^{M,N} DP_i(a,b)}{M*N} * 100 > T 
    \end{equation*}
    
    \item \textbf{Threshold-Based Detection Mode} this mode looks at the average intensity of the current frame, computed by averaging the R, G, and B values over the whole frame, and triggers a scene break when the delta in intensity between two successive frames falls below a given threshold. Zhang et. al. \cite{Zhang93} call this method \textit{Histogram comparison}. If we denote $H_i(j)$ the color histogram of frame $i$ for the channel $j$ (either R, G, B), a boundary is detected by the formula:
    \begin{equation*}  
    SD_i = \displaystyle\sum_{j=1}^{3} |H_i(j)-H_{i+1}(j)| > T 
    \end{equation*}
\end{itemize}

After a few tests, we found that the content-aware mode with the default threshold (30) separated videos into correct shots.

\subsection{Color histogram}

The point of this research is to create a music video from extracts of MVs of various origins. How does our algorithm decide which segments to select among all the different MVs ?

One first lead could be focusing on the lyrics meaning to select the MV segments, like Cai et. al. \cite{Cai07} or Wu et. al. \cite{Wu12} who create music videos from images in relation with the lyrics. However our observations and many sociological studies of MVs like Sede\~no et. al.'s \cite{Sedeno16} or the first chapter of \textit{Experiencing Music Video: Aesthetics and Cultural Context} \cite{Vernallis04} point out that song lyrics, nor a clear narrative, are not decisive elements in the conceptualisation and production of MVs.

\begin{figure}[h]
  \centering
  \includegraphics[width=\linewidth]{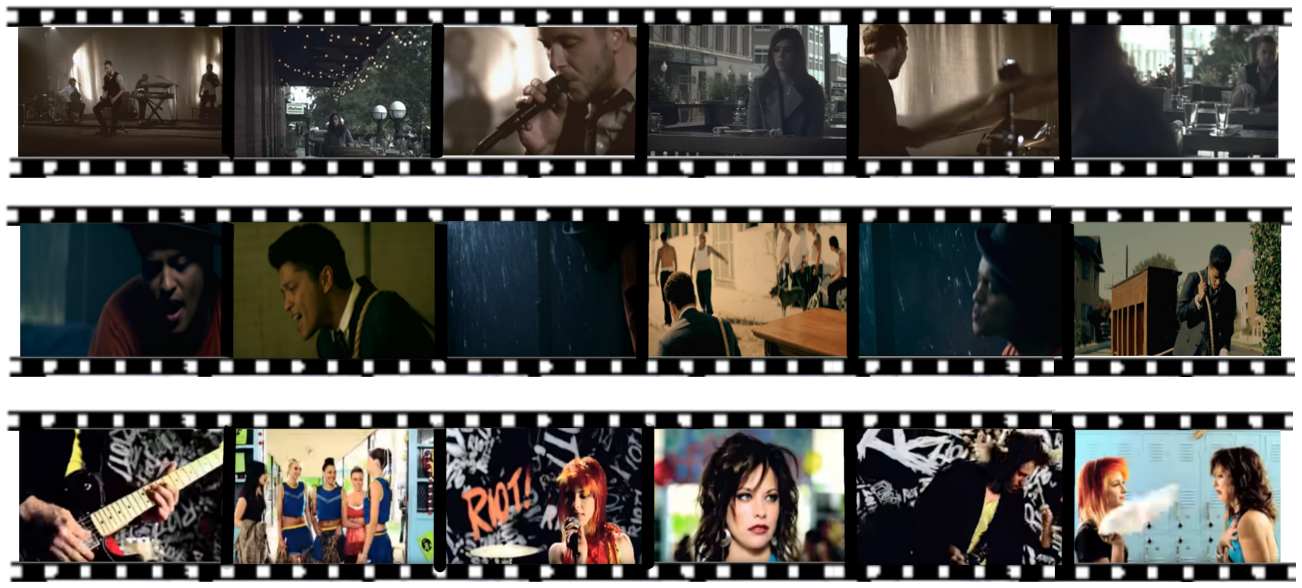}
  \caption{Alternation of styles for 3 popular MVs. From top to bottom: Secrets - One Republic, Grenade - Bruno Mars, Misery Business - Paramore.}
  \Description{Image showing for 3 music videos the alternation of styles, by showing screenshots of different moments in the MVs.}
  \label{fig:styles}
\end{figure}

Sede\~no et. al.'s research \cite{Sedeno16} identified a recurring phenomenon in MVs: western professional music videos are often organized into two "styles" which alternate throughout the video, usually one depicting the side story written in the lyrics (called \textit{concept videos}) while the other one is centered on the music performance, showing for instance the music group playing or the singer singing (called \textit{performance videos}), as illustrates figure~\ref{fig:styles}. We observed that pop songs can even have more than two "styles" to appear more dynamic.

The main feature in common for same-style scenes is the \textbf{color}. Usually a stark change is made in the color distribution of the frame in order to clearly indicate to the viewer that the video is showing content happening in a different time and spatial frame than a second ago. Thus, in order to create the same \textit{style} structure in the generated MVs, we decide to group together video shots with similar color distribution, then alternate between the clusters hence generated.

The color distribution of an image is reflected by its \textbf{color histogram}. The color histogram for a given channel represents the statistical distribution of pixels in the image for this channel. For instance if we have $N$ pixels of values $(r,g_i,b_i), 0 \leq r,g_i,b_i \leq 255$:
\begin{equation} 
    Hist_{red}[r] = N
\end{equation}

Equation (1) implies :
\begin{equation*} 
    \displaystyle\sum_{c=0}^{255} Hist_{channel}[c] = P, \quad channel \in (R,G,B)
\end{equation*}
where $P$ is the total number of pixels in the image.
As the color histogram is defined only for an image and a channel, we compute a 768-size array for representing the video color histogram :
\begin{enumerate}
    \item Extract a video frame every 5 frames (~100ms, reaction time of human eyes).
    \item Concatenate these extracted frames to get an image representing the whole video.
    \item Compute the normalized color histogram for this image on each channel (B, G, R).
    \item Concatenate the 3 arrays size 256 of each channel to get 768-size array.
    \item Store this array in a \verb+json+ file, as well as the video segment duration.
\end{enumerate}
Step 3 is executed with the help of \verb+cv2+ Python library. K-Means algorithm is executed with \verb+scikit-learn+ applied on the array stored in the \verb+json+ files.

\begin{figure}[h]
  \centering
  \includegraphics[width=\linewidth]{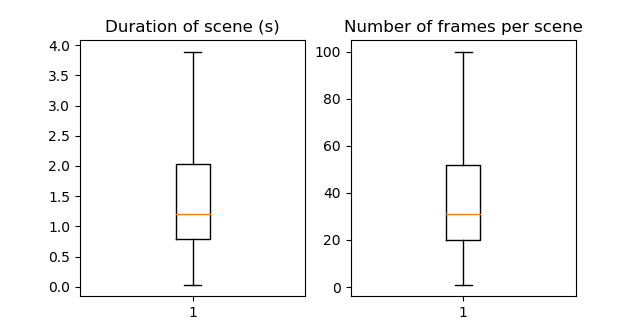}
  \caption{Box plot of database videos scenes characteristics.}
  \Description{Statistical plots of the number of scenes per video, duration of scene in seconds, number of frames per scene}
  \label{fig:stats_scenes}
\end{figure}

The computation above is executed only for scenes containing between 12 and 125 frames. Statistical analysis of our database shows that scene length varies between 0 and 100 frames, values outside these boundaries being considered as extremes (see figure~\ref{fig:stats_scenes}), hence the $[0,125]$ interval. Finally we consider scenes shorter than 12 frames (almost 500 ms) as unfit, as it is the time it takes for human eyes to process information.

\subsection{Database cleanup}
Unfortunately, many amateur MVs are incorrectly labelled as "official music video" on Youtube. As we can not afford to manually verify every video provided by \verb+Youtube-5M+ dataset, our database contains many MVs which segments are unsuitable for generating a music video. 

We first roughly remove unsuitable videos based on the video and scene length. Based on the box plots of video and scene properties, we eliminate videos containing scenes longer than 60 seconds, as this represents more than 1/4 of the average length (approximately 240s) and significantly exceeds the 1.5 interquartile range corresponding to the upper quartile of scene lengths (4s). Inspection of such videos in database shows that they are indeed usually unsuitable, for instance makeup video tutorials or lyrics videos.

\begin{figure}[h]
  \centering
  \includegraphics[width=\linewidth]{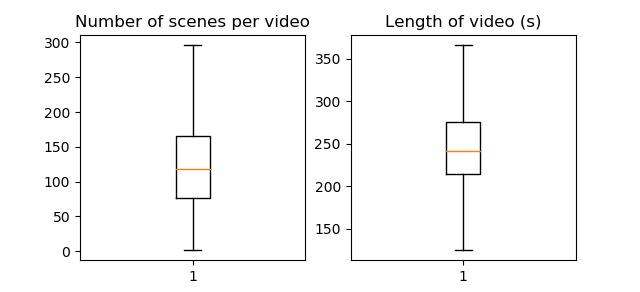}
  \caption{Box plot of database videos characteristics.}
  \Description{Statistical plots of videos length and scene number in database}
  \label{fig:stats_videos}
\end{figure}

After this step, the most common problems remaining are either MVs with hard-written lyrics, or videos made from video shots of games or movies, on which is applied a music track reflecting the desired atmosphere. In both cases, segments originating from such kind of videos clearly give off an odd impression of mismatch on the overall music video and decrease the quality of the output MV.

Another major issue are the MVs \textit{actual} resolution. Even though most videos on Youtube have a $16 \times 9$ format, the size of the black bars on top and bottom of the video have no official standard. Besides, some Youtube music videos present HD resolution ($1280 \times 720$) while others have lower quality ($640 \times 360$). The resulting MV generated from extracts of videos of different resolutions would therefore suddenly change size, which can be noticed in some media players like \verb+VLC+, or have the black bars surrounding the video constantly changing. Although most man-made amateur MVs present this default, we tackle it in order to generate better realistic-looking music video.
For this purpose, we design an algorithm to harmonize both the outer size (resolution) and the inner size (within the black bars) of the videos. First, we resize all videos to have a final resolution of $640 \times 360$. Such low resolution is chosen since more than half of videos on the database originally had this resolution, thus resizing them bigger would give very pixelated output. Besides, choosing lower sizes for the videos provides additional storage space on the testing servers. Secondly, by cropping videos, we change the black bars to fit into two categories, either no black bars at all, or an inner resolution of $640x272$. These values, again, are chosen based on statistics of the original inner resolutions of our database videos.

\section{Generation algorithm}

Our algorithm takes as input an audio file, and outputs a video using the input music as background. This algorithm is made of several steps to bring the output video as close as possible to a man-made MV:
\begin{enumerate}
    \item Find boundaries in the input music at important music changes
    \item Find input music genre
    \item Fetch in database videos with the same genre as found at step (2)
    \item Apply K-Means on the color histogram feature to cluster together scenes from videos found at step (3)
    \item Randomly select $C$ clusters from clusters created at (4)
    \item Assemble them around the boundaries found at step (1).
\end{enumerate}
Based on the lengths of videos on our database (figure~\ref{fig:stats_videos}), our algorithm rejects after step (1) inputs of length $\notin [60,400]$.
\begin{figure}[h]
  \centering
  \includegraphics[width=\linewidth]{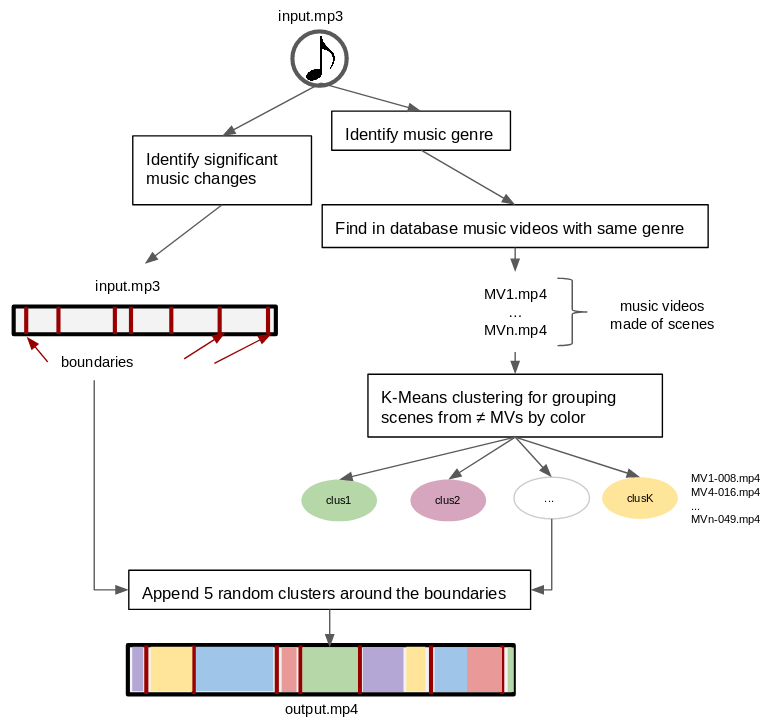}
  \caption{Flowchart of generation algorithm.}
  \Description{Generation algorithm steps}
\end{figure}
Step (2) is optional in our algorithm. If skipped, then step (4) is applied on the videos of the whole database, however the resulting output presents significantly less consistency.
We detail each step of this algorithm in the following subsections.

\subsection{Boundary detection}

First and foremost, which rhythmic indicators are \textit{relevant} for a music video ?

As Goodwin explains in his third chapter of \textit{Dancing in the distraction factory} \cite{Goodwin92}, considered as the Bible of MV analysis, rhythm in music video clips is not generally represented through the technique of cutting "on the beat", meaning that one can often observe a video shot transition without hearing a beat, or hear a beat while not observing any change in the video. Yet, he noticed that many videos mirrored the shifts of the \textit{harmonic developments} in a song. When there is a \textbf{key shift} in a melody (transition between sections such as chorus, verses, solo...), one can almost be sure to also observe a significant change in the video.

That is why we decide not to detect small rhythm changes, but only major changes between sections.
For that purpose, after testing the boundary algorithms provided by Python library \verb+MSAF+, we settle on \textbf{Ordinal Linear Discriminant Analysis (OLDA)} from McFee et. al. \cite{Mcfee14} due to its outstanding performances. This supervised learning algorithm is especially designed to split music into functional or locally homogeneous segments (\textit{e.g.}, verse or chorus).

Before applying OLDA, structural repetition features must be computed.
First, beat-related features (Mel-frequency cepstrum coefficients or chroma) are extracted from the audio sample, then a similarity matrix is computed by linking each beat to its nearest neighbors in feature space. The repetitions appear in the diagonals of the resulting matrix. To convert diagonals into horizontals, the self-similarity matrix is skewed by shifting the $i$th column down by $i$ rows.
However nearest-neighbour method can present a few errors like spurious links or skipped connections. To solve this issue, McFee et. al. apply a horizontal median filter in the skewed self-similarity matrix, resulting in a matrix $R \in \mathbb{R}^{2t\times t}$.
Finally, as only the principal components of the matrix contain the most important factors, $R$ matrix is reduced to a matrix $L \in \mathbb{R}^{d\times t}, d << 2t$ using matrix multiplications, representing \textit{latent structural repetition}.
From $L$ matrix and the audio sample, several useful features are extracted : 
\begin{itemize}
    \item mean MFCCs
    \item median chroma vectors
    \item latent MFCC repetitions
    \item latent chroma repetitions
    \item beats characteristics : indices and time-stamps (in seconds and normalized)
\end{itemize}
Repetitive features are used for genres with verse-chords structure like rock or pop, while non-repetitive features are used for other kinds of music like jazz.

OLDA algorithm is an improved version of linear discriminant analysis algorithm (FDA) developed by R. Fisher \cite{Fisher36}. 
This supervised learning algorithm takes as input a collection of labeled data $x_i \in \mathbb{R}^D$ and class labels $y_i \in \{1,2,..,C\}$. Then it tries at the same time to maximize the distance between class centroids and minimize individually the variance of each class. For that purpose, they define a linear transformation $W \in \mathbb{R}^{D \times D}$ based on the scatter matrices :
\begin{equation*} 
    A_w := \displaystyle\sum_c \displaystyle\sum_{i:y_i=C}(x_i-\mu_c)(x_i-\mu_c)^\top 
\end{equation*}
\begin{equation*} 
    A_o := \displaystyle\sum_{c<C} n_c(\mu_c-\mu_{c+})(\mu_c-\mu_{c+})^\top + n_{c+1}(\mu_{c+1}-\mu_{c+})(\mu_{c+1}-\mu_{c+})^\top 
\end{equation*}
where $\mu_c$ is the mean of class $c$ and $n_c$ is the number of examples in class $c$. $A_o$ measures the deviation of successive segments $(c,c+1)$ from their mutual centroid $\mu_{c+}$ defined as :
\begin{equation*} 
    \mu_{c+} := \frac{n_c\mu_c+n_{c+1}\mu_{c+1} }{n_c+n_{c+1}}
\end{equation*}

McFee et. al. use the same within-class scatter matrix $A_w$ as Fisher, but they improve over Fisher's between-class scatter matrix by defining $A_o$ and $\mu_{c+}$ in order to force all segments of one song to be considered of the same class during the training. They also add a $\lambda > 0$  soothing parameter to improve numeral stability in cases when $A_w$ is singular.
OLDA optimization aims at solving the following equation:
\begin{equation} 
    w := \underset{W}{\operatorname{argmax}}\text{ tr }\Big((W^\top (A_w+\lambda I)W)^{-1}W^\top A_oW\Big)
\end{equation}

\subsection{Music genre recognition}

According to Goodwin \cite{Goodwin92} and Vernallis \cite{Vernallis04}, the content of MVs are intimately related to their music genre. A \textbf{music genre} is a category that identifies a music as belonging to a set of conventions.
Based on their works and an extensive analysis of our database ($\approx$ 1000 MVs of the most popular artists on Youtube), 4 genre categories are established for our algorithm. Music videos of these 4 groups present the following visual characteristics:
\begin{description}
    \item[Pop/Indie] Close-ups on singer, dance routines, less conceptual scenes due to focus on artist
    \item[Rock/Metal/Alternative] Musical performance, concert performance, more  conceptual scenes
    \item[Hip-Hop/Rap/RnB] Sport clothing, black performers, singer close-ups, less conceptual scenes due to focus on artist, sexualized woman, street environment
    \item[Electronic/House] Dance routines, very conceptual except for DJ concert performances, representations of people partying, nature environment
\end{description}
There is no official list of all the existing music genres. These categories are established from popular tags sharing common characteristics and frequently paired together from the website \verb+Last.fm+. Some significantly different genres, such as classical music or jazz, are not represented in our database.

Using this data, we match the input music only with segments of videos from our database which music genre corresponds to the input music genre. This way, we ensure consistency between the style of the music and the video segments.

In order to identify the input music genre, several Deep neural networks trained on musics genre identification were tested \cite{Choi17}. Unfortunately, neither using the raw weights provided on \verb+github+, or fine-tuning the available algorithms gave correct results.

Therefore, fingerprinting technology is used by our algorithm to recognize the music genre.
First, the music name and artist are identified via \verb+ACR Cloud+ fingerprinting API. If the genre is not available, an additional request is sent to \verb+Last.fm+ API with the music and artist name to finally get the input music genre.
Naturally, this method is far from ideal as it works only for music popular enough to have its fingerprint recorded in \verb+ACR Cloud+ database. To remedy this problem, we also allow the user to manually input his music genre in case it is not recognized.

The above method applied to the audio channel of the videos is used to get the genre of the music videos in our database.

\subsection{K-Means on color histogram}

From the previous step, we obtain a list of music videos having the same genre as the input music. Theses music videos were previously split into scenes in the database. In order to group together different scenes from these MVs, we perform a \textbf{K-Means algorithm} using library \verb+scikit-learn+ on the color histograms previously stored in \verb+json+ files. 
This unsupervised machine learning method assigns a set of points to $K$ clusters following this process:
\begin{enumerate}
    \item Randomly position initial centroids $(c_i)_{i=1}^K$
    \item Form $K$ clusters by assigning each point $x$ to to its closest centroid:
    \[ \underset{c_i, i=1..K}{\operatorname{argmin}} dist(c_i,x) \]
    \item Recompute the centroid of each cluster of points $C_i$:
    \[ c_i = \frac{1}{|C_i|} \displaystyle\sum_{x_i \in C_i} x_i \]
    \item Repeat steps 2 and 3 until no more assignment change is observed.
\end{enumerate}

To select the $K$ value, we iterated over several value of $K$ from 10 to 100 and checked the quality of the clustering. We eventually chose $K=90$ to have a good partition of colors. 
From this step, we obtain a list of $K$ clusters made of scenes originating from MVs with the same genre as the input music.

\subsection{Final build}
Given the boundaries array of the music segmentation and a set of color clusters, we can finally assemble the final music video:
\begin{enumerate}
    \item Randomly select $C$ clusters large enough to cover whole song.
    \begin{equation*} 
    \displaystyle\sum_{k=0}^{C} \displaystyle\sum_{i} l^k_i > l_{input} , \quad l^k_i=\text{length of scene $i$ in cluster $k$}
    \end{equation*}
    \item Shuffle scenes grouped by MV in cluster.
    \item As long as boundary not met, append videos of same cluster. If meet end of cluster, start appending next cluster.
    \item When meeting a boundary, switch to next cluster.
    \item Repeat steps 3 and 4 until (end of the song - {\tiny END\_OFFSET}) sec.
    \item Try find one last scene to cover the whole end. If find, append it and add a fading effect.
\end{enumerate}

We tried several values of $C$ before settling to value $C=5$. The last step was added later after noticing that the end of the generated music videos looked odd. After comparing to real MVS, we noticed the reason was because our generated MVs still had frequent changes of scenes at the end of the video, while in real MVs the length of scenes grew bigger at the end, with less scene changes.

Step 2 was added so that if two generated MVs use the same cluster, the scenes issued from different original MVs would not appear in the same order, making these 2 MVs look different.

\section{Experiments}

\subsection{Experimental configuration}
The original goal of this research was: can an AI algorithm create music videos comparable to human-made MVs?

Human-made MVs can be separated into two groups: \textit{professional MVs} and \textit{amateur MVs}. Professional MVs are created by a production team with professional cameramen, editors, etc. On the other hand, amateur MVs are music videos created by people as a hobby, usually because they are fans of the music or the artist. Either they shoot themselves a whole new music video, or they assemble extracts of previous music videos from that artist via a video editing software to create a realistic-looking music video for that song. We will consider the latter kind of amateur MVs for this study.

We asked volunteers to judge a total of 30 music videos and classify them into one of the three categories: generated, professional of amateur MV. They should as well explain in a text input field why they made this decision.
In these 30 MVS :
\begin{itemize}
    \item 15 were generated music videos, selected for their quality
    \item 7 were professional music videos, selected randomly on Youtube
    \item 8 were amateur music videos, selected on Youtube using 'unofficial music video' keywords
\end{itemize}

The videos used for this experiment are available at the url \url{https://sites.google.com/view/music-video-generation}.

We used Amazon Mechanical Turk to assess each MV to exactly 10 users. This website is a crowdfunding platform enabling people to perform remunerated tasks. For our batch of 30 videos, each worker could complete as many tasks as he could before all the tasks are completed. Workers needed a 90\% record of correct evaluations on MTurk to be accepted to perform this experiment

\subsection{Experimental results and analysis}
In total, 126 different workers classified our $30 \times 10$ tasks. We controlled their classification labels based on the number of videos each worker classified, the correctness of their guesses or the detail of their justifications for their choices, to eliminate results from workers who seemed to give random answers. In order not to influence the worker's decisions, we explained each category with the following terms :
\begin{description}
    \item[generated] MV created by an AI algorithm
    \item[professional]  official MV created for the artist by a professional production team
    \item[amateur] fan-made MV
\end{description}

\begin{figure}[h]
  \centering
  \includegraphics[width=\linewidth]{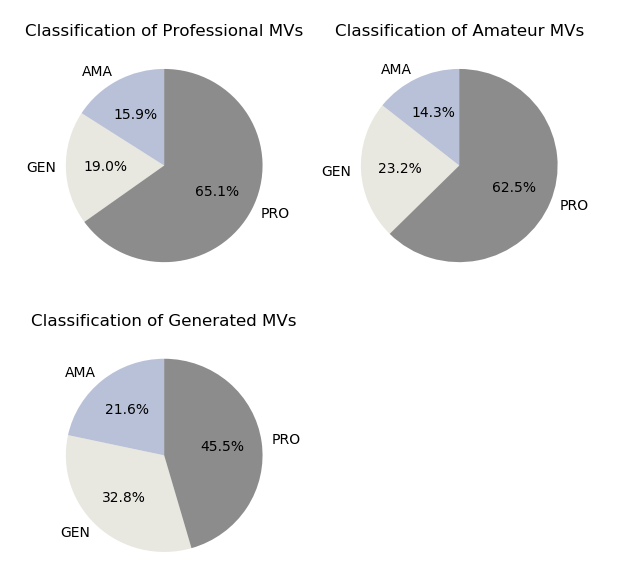}
  \caption{Percentage of answers for each category of music video. Light grey: generated MV answers, middle grey: amateur MV answers, dark grey: professional MV answers.}
  \Description{Three pie charts showing the percentage of answers for each category of music video. Results are the following. For professional MVs, 65.1\% were correctly classified as professional, 15.9\% were classified as amateur, and 19\% were classified as generated. For amateur MVs, 14.3 \% were correctly classified as amateur, 62.5\% were classified as professional, and 23.3\% were classified as generated. Finally for generated MVs, 32.8\% were correctly classified as generated, 45.5\% were classified as professional, and 21.6\% were classified as amateur.}
  \label{fig:results_classification}
\end{figure}

We grouped together classifications answers for each category of music video to evaluate the performance of our algorithm on figure~\ref{fig:results_classification}.

Results show that \textbf{our generated MVs are most often perceived like human-made videos}. 45\% of generated videos were mistaken for professional music videos, and 21.6\% were mistaken for amateur-made music videos. Users seem to have little knowledge of amateur MVs, as they mostly classified them as professional videos. Nonetheless these results show logic through the following points:
\begin{itemize}
    \item The percentage of classification as "generated" increases with the non-professionalism of the video : they represent 19\% of pro MVs, 23.2\% of amateur MVs, and 32.8\% of generated MVs.
    \item The percentage of classification as "professional" decreases with the non-professionalism of the video : they represent 65.1\% of pro MVs, 62.5\% of amateur MVs, and 45.5\% of generated MVs.
    \item The percentage of classification as "amateur" is the greatest for generated videos, which mean they were confused for human-made videos but the users still noticed some odd characteristics preventing them to give the "professional" label.
\end{itemize}

By grouping together amateur MVs and professional MVs in one category named human-made MVs and considering "generated" as positives and "human-made" a negatives, we can evaluate classification metrics for this experiment in tab~\ref{tab:metrics}.
\begin{table}
  \caption{Classification metricss}
  \label{tab:metrics}
  \begin{tabular}{ccl}
    \toprule
    Metric & Value & Interpretation\\
    \midrule
    accuracy & $0.55$ & how correctly the users classified\\
    precision & $0.64$ & how accurate are the "generated" predictions\\
    sensitivity & $0.33$ & how well they recognized generated videos\\
    specificity & $0.79$ & how well they recognized man-made videos\\
  \bottomrule
\end{tabular}
\end{table}

From these metrics, we can interpret that users do not recognize generated videos when they see one in 2/3 of cases, but when they give the label they are mostly correct (64\%). They  are reluctant to giving the "generated" label even when they see a generated video, which is why we have a high specificity but mediocre accuracy.

\begin{figure}[h]
  \centering
  \includegraphics[width=0.9\linewidth]{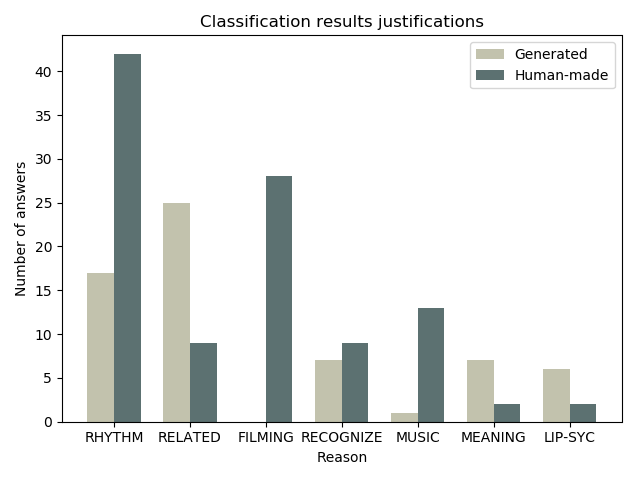}
  \caption{Reasons given by the workers to justify their classification. }
  \Description{Bar chart representing, for 7 different reasons extracted from the workers' answers, how many labels "generated" and how many labels "human-made" were given based on each reason.}
  \label{fig:results_reasons}
\end{figure}

In order to understand better the choices of the workers, we analyze the reasons they gave while they classified. The number of resulting labels for a reason are shown in figure~\ref{fig:results_reasons}.

The main reasons quoted by workers are :
\begin{description}
    \item[RHYTHM] : how the music and video were in synchronization.
    \item[RELATED] : how the video shots presented coherence altogether. 
    \item[FILMING] : quality of the filming and video effects. 
    \item[RECOGNIZE] : the user recognized either artists and pieces in the video, or the author of the music, hence helping him to decide. 
    \item[MUSIC] : quality of the music in the video.
    \item[MEANING] : whether the video content were in relation with the lyrics.
    \item[LIP-SYNC] : whether lip-syncing was correctly performed.
\end{description}
A few elements better explain why such of big proportion of videos were classified as either "professional" or "amateur". First, since we decided to base our algorithm not on videos generated from scratch but from extracts of real videos, the video outcome always use segments of videos with professional filming, sometimes presenting video effects. This is why no videos in the whole experience were classified as "generated" when users judged based on the video quality. This reason also explain confusions between amateur and professional videos. The amateur mashup of high-resolution videos from Chris Brown clips was unanimously categorized as "professional", although a knowledgeable viewer could easily detect the amateur provenance of this video through the black bars and the watermarks. On the other hand, the official MVs \textit{Believe} from Cher or \textit{Get Thru This} from Art of Dying were frequently mistakenly classified as "amateur" or "generated" due to poor resolution quality. Bold artistic choices in the category "relatedness" can also induce classification errors: the official MV \textit{Money} from Pink Floyd received the most "generated" labels in the whole batch, due to recurrence of shots representing coins and the apparent randomness of other shots. However, knowing the name of the music and paying attention to the lyrics, one can understand that the "random" and coin shots actually represent a critic of American consumerism. In opposite, amateur and generated music videos used more conventional shots, such as concert extracts, people partying, etc. Yet, the "randomness" of video shots, meaning the lack of coherence, appear to be a correct lead to identify generated videos, as this was the main reason given for our generated MVs.
Second, the quality of the music is one factor helping the workers to decide whether the video is generated. Since our algorithm takes in input a real music instead of generating it, naturally the workers would tend to give a "human-made" label if they judge based on this criterion.
Finally, as predicted in Section 3, the meaning of lyrics represent only a small proportion of the reasons given by the workers to evaluate a music video. 

To further interpret these results, we show which proportion of these labels actually come from generated videos, in order to know the performance of our algorithm in each of the reason categories shown above. Since we aim at generating realistic-looking videos, receiving the label "generated" with a given reason category means the generated video performed badly in this category, while a "human-made" label means the generated video performed well enough in this category. 

\begin{figure}[h]
  \centering
  \includegraphics[width=0.9\linewidth]{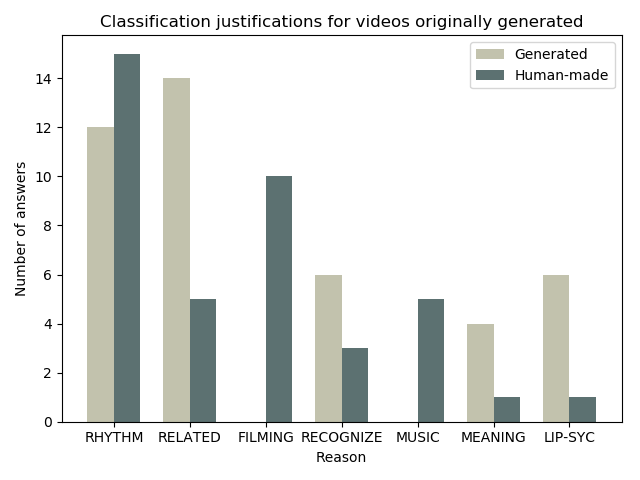}
  \caption{Justifications given by the workers when they were actually classifying a generated MV. }
  \Description{Bar chart representing, for 7 different reasons extracted from the workers' answers, how many labels "generated" and how many labels "human-made" were given based on each reason, when the videos seen were actially music videos.}
  \label{fig:results_reasons_gen}
\end{figure}

In 55.6\% of the cases, rhythm was judged well-performed enough to induce workers in classifying the generated video as "human-made". Thus, main leads to improve are the relatedness of the video shots and the lip-syncing.
 
\section{Conclusion and future work}
In this paper, we proposed a novel technique for generating a music video from extracts of Youtube music videos. This technique was based of sociological observations of MV structures. The results of the user study show that generated MVs could easily be mistaken for amateur and even professional music videos. Users feedback show that the quality of video shooting, how the shots make a coherent ensemble and synchronization with the rhythm are the most important elements to easily recognize a realistic-looking music video.

As our algorithm only takes about 1m30s to run, due to the storage in advance of features and K-Means results, we made it publicly testable on a website, where people can generate their own video clips for their music tracks. After inputting their audio file, users are informed of the progression of the generation algorithm using a modal, and can download the video after a little wait. This website received high interest as over 300 videos have been generated until now. 

There are still many opportunities in improving this algorithm. In 100 generated MVs, half of them needed to be discarded because they contained a segment of an unsuitable MV (lyrics video, piece of cartoon, etc.) which immediately could inform the viewer of the nature of the MV. In these 100, 10 had good quality enough to be used in the experimentation.

Thus, a method for automatically cleaning the database would significantly improve our algorithm's performance, as in this case 20\% of the generated videos could be considered of "good quality" instead of 10\%. For detection of lyrics music videos, one could implement a machine learning algorithm for recognizing text in the video frames. For detection of cartoons, one could perform further analysis in the distribution of colors in the frames to detect the percentage of flat color sections.

A further step would be to improve the video-music synchronization and the illusion of lip-syncing by matching music segments with no singing only with video segments where no mouth is moving, and only using mouth-opening video segments of people with the same gender as the input music singer.

Finally, further research on genre recognition would provide a new method more resilient to input music variety.

Participation to this project can be done via its public \verb+github+ repository.

%

%
\bibliographystyle{ACM-Reference-Format}
\bibliography{sample-base}

%

\end{document}